\documentclass[11pt, a4paper]{article}
\usepackage[absolute]{textpos}
\usepackage[top=3cm,bottom=3cm,left=2cm,right=2cm]{geometry}
\usepackage{amsmath,amssymb,mathtools,dsfont,emptypage,graphicx,dcolumn,bm,booktabs,colortbl,collcell,enumerate,float,verbatim,multirow,xparse,slashed,mathrsfs,color}
\usepackage{cite}
\usepackage{braket}
\usepackage{tabularx}
\usepackage{multirow}
\usepackage{graphicx}
\usepackage{subcaption}
\usepackage{appendix}
\usepackage{xcolor}
\usepackage{pifont}
\usepackage[labelfont=bf, labelsep=period]{caption}
\usepackage{multirow}
\usepackage{here}
\usepackage{physics}
\usepackage[T1]{fontenc}
\usepackage{siunitx}

\usepackage{float}

\usepackage[compat=1.1.0]{tikz-feynman}
\usepackage{contour}%FeynamnDiagrams e contorni

\usepackage[colorlinks=true,urlcolor=blue,linkcolor=black,citecolor=blue]{hyperref}

\begin{document}

\begin{center}
{\bf\Large Minimal Majoron Dark Matter} \\
[5mm]
\renewcommand*{\thefootnote}{\fnsymbol{footnote}}
Kensuke Akita$^{a}$\footnote{\href{kensuke@hep-th.phys.s.u-tokyo.ac.jp}{\tt kensuke@hep-th.phys.s.u-tokyo.ac.jp}},
Koichi Hamaguchi$^{a,b}$\footnote{\href{mailto:hama@hep-th.phys.s.u-tokyo.ac.jp}{\tt
 hama@hep-th.phys.s.u-tokyo.ac.jp}}, Haruto Kitagawa$^{a}$\footnote{\href{kitagawa@hep-th.phys.s.u-tokyo.ac.jp}{\tt kitagawa@hep-th.phys.s.u-tokyo.ac.jp}} and Tatsuya Yokoyama$^{a}$\footnote{\href{mailto:tyokoyama@hep-th.phys.s.u-tokyo.ac.jp}{\tt tyokoyama@hep-th.phys.s.u-tokyo.ac.jp}}\\
\vspace{2mm}
$^{a}$\,{\it Department of Physics, University of Tokyo, Bunkyo-ku, Tokyo 113--0033, Japan,}\\
 $^{b}$\,{\it Kavli Institute for the Physics and Mathematics of the Universe (Kavli IPMU), University of Tokyo, Kashiwa 277--8583, Japan,}
\end{center}

\begin{abstract}
We study Majoron dark matter (DM) in its minimal realization, based on the Type-I seesaw framework extended by a SM-singlet complex scalar. 
Remaining agnostic about the origin and value of the Majoron mass, 
we evaluate the DM abundance from both the freeze-in and misalignment mechanisms, and identify the viable parameter space consistent with observational constraints. 
Without fine-tuning of the initial misalignment angle, we find that the Majoron mass is bounded by $m_J \lesssim \mathcal{O}(10)~\mathrm{MeV}$.
We also discuss compatibility with thermal leptogenesis. 
Successful leptogenesis with two right-handed neutrinos favors misalignment-dominated production with the Majoron mass $m_J \lesssim \mathcal{O}(100)~\mathrm{eV}$, while freeze-in dominated production is compatible with leptogenesis only with a mild fine-tuning 
of the initial misalignment angle, $\theta_i \lesssim \mathcal{O}(0.01)$.
\end{abstract}

\tableofcontents

\renewcommand*{\thefootnote}{\arabic{footnote}}
\setcounter{footnote}{0}

\section{Introduction}

The origin of the tiny neutrino masses remains an outstanding puzzle in particle physics, pointing to physics beyond the Standard Model (SM). Arguably, the most economical and well-known solution to this problem is the so-called Type-I seesaw mechanism \cite{Minkowski:1977sc,Yanagida:1979as,Gell-Mann:1979vob,Gell-Mann:1979ijt,Mohapatra:1979ia,Schechter:1980gr}. In its minimal realization, we introduce right-handed neutrinos (RHN) with heavy Majorana masses, typically far above the electroweak scale. Then, these heavy RHN Majorana masses suppress the masses of the SM neutrinos.
Interestingly, RHN may also serve as a key to another problem, the Baryon Asymmetry of the Universe (BAU), via leptogenesis\cite{Fukugita:1986hr}. The RHN Majorana mass terms violate $\mathrm{U(1)_{B-L}}$, and their decays provide the origin of the asymmetry. In both the seesaw mechanism and leptogenesis, the RHN Majorana mass terms play a crucial role. If these masses are generated by the spontaneous symmetry breaking (SSB) of a global $\mathrm{U(1)_{B-L}}$, a (pseudo) Nambu-Goldstone (NG) boson, known as the Majoron, arises~\cite{Chikashige:1980ui, Gelmini:1980re, PhysRevD.25.774}.

The Majoron would be a massless particle if the global $\mathrm{U(1)_{B-L}}$ were an exact symmetry. However, it is widely expected that global symmetries are violated by quantum-gravity effects~\cite{Giddings:1989bq, ABBOTT1989687}.
Thus, the Majoron can acquire a mass due to the explicit breaking of the global $\mathrm{U(1)_{B-L}}$ induced by high-energy physics. The interactions of the Majoron with other fields are suppressed by the vacuum expectation value (VEV) of the scalar field $\sigma$ responsible for the SSB of $\mathrm{U(1)_{B-L}}$, which makes the Majoron long-lived and weakly interacting. 
In such a situation,
the Majoron can be a candidate for the dark matter (DM) in the Universe. 
The Majoron framework therefore provides a potentially economical setup that can simultaneously account for tiny neutrino masses, BAU, and DM.

In this paper, we consider Majoron DM in its minimal realization. We work within the Type-I seesaw framework with two RHNs, extended by a SM-singlet complex scalar $\sigma$ whose VEV induces the SSB of $\mathrm{U(1)_{B-L}}$. We remain agnostic about the origin of the Majoron mass, treating it as a phenomenological parameter, 
and assume that the Majoron interactions are not affected by the explicit $\mathrm{U(1)_{B-L}}$ breaking.
We study Majoron DM production from both the freeze-in mechanism \cite{Hall:2009bx} and the misalignment mechanism. 
We evaluate the Majoron DM abundance and present current experimental constraints and future prospects. 
We also discuss the compatibility with thermal leptogenesis and show that 
successful leptogenesis with two RHNs favors 
misalignment-dominated production, while freeze-in dominated production 
requires a mild fine-tuning of the initial misalignment angle.

The thermal production of Majoron DM has been extensively studied in the literature \cite{Rothstein:1992rh,Berezinsky:1993fm, Lattanzi:2007ux, Bazzocchi:2008fh, Gu:2010ys, Lattanzi:2013uza, Queiroz:2014yna} within the freeze-out scenario. 
The freeze-in production~\cite{Frigerio:2011in, Boulebnane:2017fxw, Heeck:2017xbu,Abe:2020dut, Manna:2022gwn,Xu:2023xva,King:2024idj,Greljo:2025suh,Chiang:2026ayw} and the misalignment production~\cite{Greljo:2025suh} of Majoron(-like) DM have been studied, although the model and/or production processes are different from those in our work.
In this work, we consider both the freeze-in production and the misalignment production of Majoron DM in its minimal realization. In addition, we also discuss the compatibility with thermal leptogenesis,
examining the effects 
of Yukawa structure and RHN mass hierarchies on the viable parameter space.
There also exist other mechanisms for the production of Majoron DM via topological defects \cite{Reig:2019sok} and kinetic misalignment \cite{Chun:2023eqc,Berbig:2025hlc}.
Constraints on Majoron DM from observations such as cosmic neutrinos, cosmic-rays,  X- and gamma-rays, and CMB have been analyzed in Refs.~\cite{Lattanzi:2007ux,Bazzocchi:2008fh,Lattanzi:2013uza,Garcia-Cely:2017oco, Akita:2023qiz}. 
Majorons in related models may also be probed through axion detection experiments 
and gravitational wave observations~\cite{Liang:2024vnd, Obata:2026qwx}.

The paper is organized as follows. In Sec.~\ref{sec:setup}, we describe our model, based on the Type-I seesaw Lagrangian extended with a complex scalar. In Sec.~\ref{sec:production}, we discuss the production of Majoron DM and present current experimental constraints as well as future sensitivities. In Sec.~\ref{sec:Majoron_DM_and_LG}, we examine the implications of the Majoron model for thermal leptogenesis. Finally, we conclude in Sec.~\ref{sec:conclusion}.
In Appendix~\ref{sec:appendix_general_case}, 
we discuss how the results are modified for nondegenerate RHN masses and nonzero Casas-Ibarra parameters.

\section{Model setup}
\label{sec:setup}

We consider a minimal extension of the Standard Model with a global $\mathrm{U(1)_{B-L}}$ symmetry and RHNs, described by the Lagrangian\footnote{This setup generically gives rise to corrections to the Higgs mass parameter, from both the tree-level coupling $|\sigma|^2|H|^2$ and loop effects. We do not address the possible fine-tuning issue in the present work.
}
\begin{align}
    \mathcal{L}&=\mathcal{L}_{\mathrm{SM}}+i\bar{N}_i\slashed{\partial}N_i+|\partial_\mu\sigma|^2
    -V(\sigma)
    -\left[\frac{1}{2}g_i\sigma \bar{N_i}^cN_i+y_{\alpha i}\bar{L}_\alpha \tilde{H}N_i+\mathrm{h.c.}
    \right]. \label{U1_B_L_lagrangian}
\end{align}
Here, $H$ and $L_\alpha$ ($\alpha = e, \mu, \tau$)
are the SM Higgs doublet and the lepton doublets, respectively, and $\tilde{H}=i\sigma_2 H^*$.
We focus on the minimal case with two generations of RHNs, $N_i$ ($i=1,2$).
The field $\sigma$ is a SM-singlet complex scalar with $B-L$ charge $2$.

We assume that the scalar potential is given by
\begin{align}
    V(\sigma) =\lambda \left(|\sigma|^2-\frac{f^2}{2}\right)^2, \label{sigma_potential}
\end{align}
which spontaneously breaks the $\mathrm{U(1)_{B-L}}$ symmetry.
After the SSB of $\mathrm{U(1)_{B-L}}$, $\sigma$ obtains a non-zero VEV $\langle\sigma\rangle=\frac{f}{\sqrt{2}}$, and we parametrize $\sigma$ around its VEV by
\begin{align}
    \sigma&=(f+\rho+iJ)/\sqrt{2}. \label{Majoron_def}
\end{align}
Here, $\rho$ is a real scalar, and $J$ is the NG boson, which is the Majoron. 
We take $f$ to be real and positive without loss of generality.

After the $\mathrm{U(1)_{B-L}}$ symmetry breaking, the RHNs acquire Majorana masses and interact with $\rho$ and $J$ as
\begin{align}
    \mathcal{L}\supset& -\frac{1}{2}M_i\bar{N}^c_iN_i-\frac{g_i}{2\sqrt{2}}\rho \bar{N}^c_iN_i-\frac{i}{2\sqrt{2}}g_i J\bar{N}^c_i N_i+\mathrm{h.c.},\\
    M_i&=\frac{g_i}{\sqrt{2}}f.
\end{align}
We have chosen our basis to diagonalize the mass matrix of the RHNs, and assume $g_2\geq g_1$. 
The radial mode $\rho$ has a mass given by $m_\rho=\sqrt{2\lambda}\,f$.
In the following, we assume $T_{\rm{R}} < f$
and $T_{\rm{R}} < m_\rho$, where $T_{\rm{R}}$ is the reheating temperature after inflation, so that the $\mathrm{U(1)_{B-L}}$ symmetry is not restored and $\rho$ is not thermally produced after reheating.

After the SSB of the SM electroweak symmetry, the SM neutrinos obtain their tiny masses through the Type-I seesaw mechanism~\cite{Minkowski:1977sc,Yanagida:1979as,Gell-Mann:1979vob,Mohapatra:1979ia,Schechter:1980gr}:
\begin{align}
    m^\nu_{\alpha\beta}=-
    \left(m_D M^{-1}m_D^T\right)_{\alpha\beta},
\end{align}
where ${m_D}_{\alpha i}=y_{\alpha i}v_{\mathrm{EW}}/\sqrt{2}$
with $v_{\rm EW}\simeq 246~\mathrm{GeV}$. In this framework, it is convenient
to use the Casas-Ibarra parameterization \cite{Casas:2001sr}
\begin{align}
m_D&=iU\sqrt{\hat{m}^\nu}R^T\sqrt{M},
\label{eq:CI-para}
\end{align}
to express the Yukawa coupling $y$ in terms of the Pontecorvo-Maki-Nakagawa-Sakata (PMNS) matrix $U$~\cite{ParticleDataGroup:2024cfk}, the diagonalized mass of the SM neutrino $\hat{m}^\nu$, and the RHN masses.
Since we assume two RHNs, the lightest SM neutrino is massless. 
For simplicity, we assume the normal ordering for the light neutrino masses in the present work. 
Accordingly, $\hat{m}^\nu_1=0$.
The matrix $R$ can be parameterized as
\begin{align}
    R=\begin{pmatrix}
        0 & \cos z & \sin z\\
        0 & -\sin z & \cos z
    \end{pmatrix}, \qquad z=a+ib\in \mathbb{C}.
\end{align}
In the numerical analysis, we use the best-fit values of the neutrino oscillation parameters for the normal ordering from NuFit 6.0~\cite{Esteban:2024eli}:  $\theta_{12}=33.68^\circ$, $\theta_{23}=43.3^\circ$, $\theta_{13}=8.56^\circ$, $\delta_{\text{CP}}=212^\circ$, $\Delta m_{21}^2=\SI{7.49e-5}{\eV^2}$, and $\Delta m_{31}^2=\SI{2.513e-3}{\eV^2}$.

So far, we have assumed that the global $\mathrm{U(1)_{B-L}}$ symmetry is exact and the Majoron is a massless NG boson. 
However, it is widely believed that global symmetries are violated by quantum-gravity effects~\cite{Giddings:1989bq, ABBOTT1989687}.\footnote{Another explicit $\mathrm{U(1)_{B-L}}$ breaking scenario is discussed in 
Ref.~\cite{Frigerio:2011in}.} 
Therefore, we allow for a small explicit breaking of $\mathrm{U(1)_{B-L}}$ that generates a mass for the Majoron.  In the following, we parameterize this effect by introducing a phenomenological mass term,
\begin{align}
\mathcal{L} \supset -\frac{1}{2} m_J^2 J^2.
\end{align}
We remain agnostic about the origin of the Majoron mass and treat it as an independent parameter of the effective theory.

\section{Minimal Majoron dark matter}
\label{sec:production}

In this section, we discuss the phenomenology of Majoron DM in the minimal setup introduced above. 
We first discuss the Majoron abundance from freeze-in and misalignment productions in Sec.~\ref{subsec:Freeze-in} and \ref{subsec:Majoron_DM_misalignment}, respectively, and then investigate current experimental constraints and future prospects in Sec.~\ref{subsec:constraints_and_prospects}.

%%%%%%%
\subsection{Majoron dark matter via Freeze-in production}
\label{subsec:Freeze-in}

In this subsection, we discuss the freeze-in production of Majoron DM. The Majoron is produced via the scattering processes 
$NN\rightarrow JJ$, $NH\rightarrow JL$, $NL\rightarrow JH$, and $HL\rightarrow NJ$.
The evolution of the Majoron number density is governed by the Boltzmann equation
\begin{align}
\frac{dn_J}{dt}+3Hn_J &= C_J,
\end{align}
where $H$ is the Hubble parameter and $C_J$ denotes the collision term for Majoron production.
Here and in what follows, $n_X$ denotes the number density of the particle $X$.
Assuming the radiation-dominated Universe, the Hubble parameter is given by
$H(T)=(\pi^2 g_*/90)^{1/2} T^2/M_{\rm Pl}$,
where $T$ is the temperature, $M_{\rm Pl}\simeq 2.4\times 10^{18}~{\rm GeV}$ is the reduced Planck scale, and $g_*=106.75$ is the effective number of relativistic degrees of freedom of the SM.

In the freeze-in regime, the Majoron interactions are sufficiently weak that the inverse processes involving Majorons can be neglected. We will later comment on the parameter region in which this approximation may break down and the Majoron can approach thermal equilibrium. 
The collision term $C_J$ is then given by
\begin{align}
C_J &= 
\sum_i \langle \sigma_{N_i N_i\rightarrow JJ} v \rangle 
n_{N_i} n_{N_i}
+\sum_{i,\alpha}\langle \sigma_{N_i H\rightarrow J L_\alpha} v \rangle 
n_{N_i} n_H^{\rm eq}
\nonumber\\
&+\sum_{i,\alpha}\langle \sigma_{N_i L_\alpha\rightarrow J H} v \rangle
n_{N_i} n_{L_\alpha}^{\rm eq}
+\sum_{i,\alpha}\langle \sigma_{H L_\alpha\rightarrow N_i J} v \rangle
n_H^{\rm eq} n_{L_\alpha}^{\rm eq},
\end{align}
where $\langle \sigma v \rangle$ denotes the thermal average of the cross section times the relative velocity.
We assume that the SM particles $H$ and $L_\alpha$ are in thermal equilibrium, so that their number densities are given by $n_H^{\rm eq}$ and $n_{L_\alpha}^{\rm eq}$, respectively. In the numerical calculations, we adopt the Maxwell-Boltzmann distributions for $\langle\sigma v\rangle$ and $n^{\rm{eq}}$, for simplicity.
By contrast, the RHNs are not necessarily thermalized, and their number densities $n_{N_i}$ are determined from a separate Boltzmann equation:
\begin{align}
\frac{dn_{N_i}}{dt} +3Hn_{N_i} &= -\langle\Gamma_i\rangle (n_{N_i}-n_{N_i}^{\rm eq})
\end{align}
where $\langle\Gamma_i\rangle$ is the thermally averaged decay rate of the RHN $N_i$~\cite{Buchmuller:2004nz}, and $\Gamma_i=(y^\dagger y)_{ii}M_i/8\pi$ is its decay width at rest.
We neglect the scattering contribution to the RHN production, for simplicity.

The relevant scattering cross sections are given by\footnote{Eq.~\eqref{eq:NN2JJ} agrees with the corresponding formula in Ref.\cite{Frigerio:2011in}.}
\footnote{While Eq.~\eqref{eq:NH2JL} diverges formally at $x_i=1$, $\langle\sigma v\rangle$ nevertheless converges.}
\begin{align}
    \sigma_{N_i N_i\rightarrow JJ}
    &=\frac{g_i^4}{256\pi}
    \frac{1}{M_i^2}
    \left(
    \frac{1}{\beta_i}
    -
    \frac{2}{x_i \beta_i^2}
    \log\frac{1+\beta_i}{1-\beta_i}
    \right)
    \label{eq:NN2JJ}
    \\
    \sigma_{N_i H\rightarrow JL_\alpha}
    &=\frac{g_i^2|y_{\alpha i}|^2}{64\pi}
    \frac{1}{M_i^2}
    \frac{x_i}{(x_i-1)^2}\ln x_i,
    \label{eq:NH2JL}
    \\
    \sigma_{N_i L_\alpha\rightarrow JH}
    &=
    \frac{g_i^2|y_{\alpha i}|^2}{128\pi }
    \frac{1}{M_i^2}
    \left(
    -\frac{1}{x_i-1}
    +
    \frac{x_i}{(x_i-1)^2}\ln x_i
    \right),
    \label{eq:NL2JH}
    \\
    \sigma_{HL_\alpha\rightarrow N_i J}
    &=\frac{g_i^2|y_{\alpha i}|^2}{128\pi}
    \frac{1}{M_i^2}
    \frac{x_i-1}{x_i^2},
    \label{eq:HL2NJ}
\end{align}
where $x_i=s/M_i^2$ and 
$\beta_i=(1-4M_i^2/s)^{1/2}$ with $s$ being the Mandelstam variable.
We have neglected the Majoron mass, assuming $s \gg m_J^2$.

In the numerical analysis in this section, for simplicity, we restrict ourselves to the case of
\begin{align}
M_1=M_2\equiv M,\qquad
z=0,
\label{eq:M1_M2_z}
\end{align}
and correspondingly set $g_1=g_2\equiv g$,
so that the Yukawa couplings are fixed solely by the low-energy neutrino parameters and the RHN mass scale $M$. 
Cases with $M_1\ne M_2$ and $z\ne 0$ are discussed in Appendix~\ref{sec:appendix_general_case} and
Sec.~\ref{sec:Majoron_DM_and_LG}.
We also assume instantaneous reheating and negligible initial abundances for both the Majoron and the RHNs. 
We fix the reheating temperature to $T_{\rm R}=10\,M$, for which the RHNs are thermalized before $T\sim M$.
We have also checked that the evolution of the RHN abundance is only weakly sensitive to $T_{\rm R}$ as long as $T_{\rm R}\gg M$.

\begin{figure}[t]
    \centering
    \includegraphics[width=0.48\linewidth]{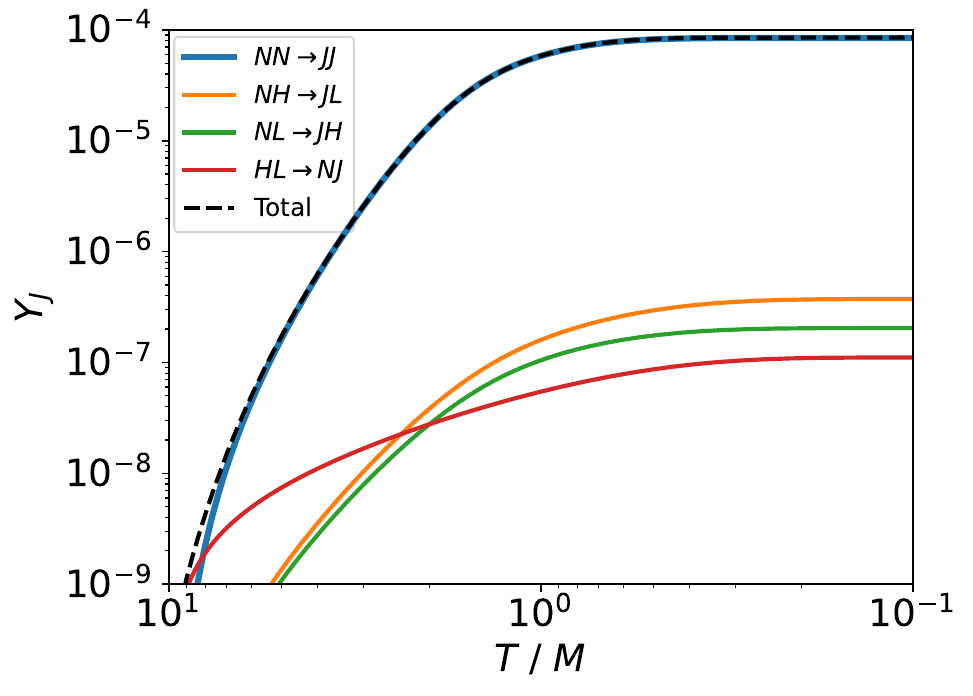}
    \includegraphics[width=0.48\linewidth]{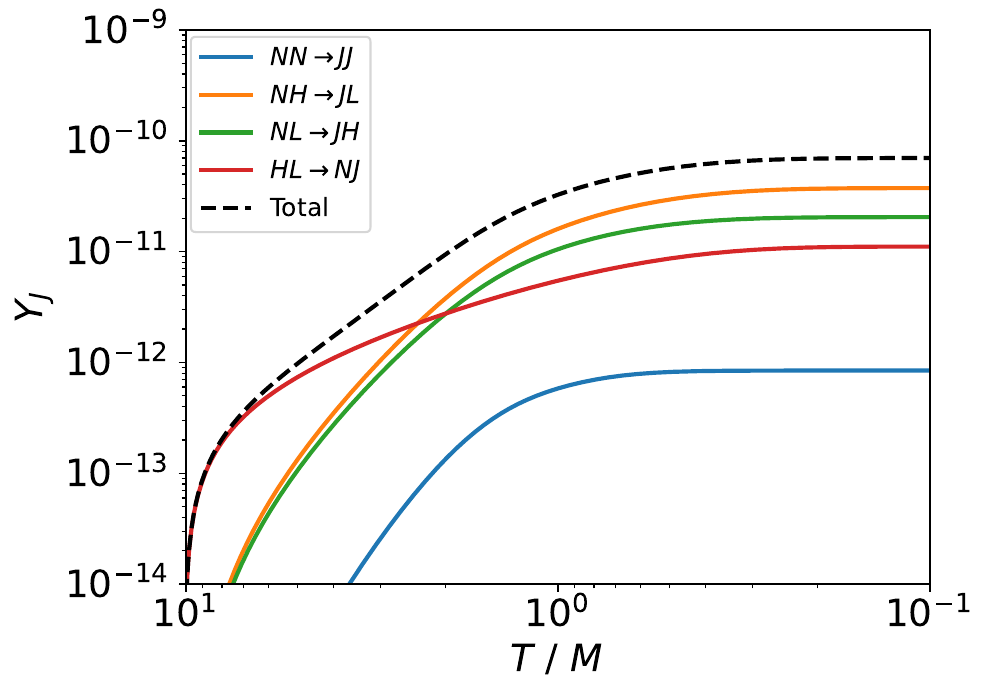}
    \caption{Evolution of the Majoron abundance $Y_J=n_J/s$ from freeze-in production for $M_1=M_2=10^8~\mathrm{GeV}$, with $g=10^{-2}$ ($f\simeq1.4\times10^{10}~\mathrm{GeV}$) in the left panel and $g=10^{-4}$ ($f\simeq1.4\times10^{12}~\mathrm{GeV}$) in the right panel. The individual contributions from the different freeze-in production channels, as well as their sum, are shown separately.
    }
    \label{fig:Yevolve}
\end{figure}

Fig.~\ref{fig:Yevolve} shows the evolution of the 
Majoron abundance from freeze-in production, $Y_J=n_J/s$, where $s=(2\pi^2/45)g_* T^3$ is the entropy density.
We take 
$M_1=M_2=10^8~\mathrm{GeV}$,
and $g=10^{-2}$ (left) and 
$g=10^{-4}$ (right), corresponding to $f\simeq1.4\times10^{10}~\mathrm{GeV}$ and  $f\simeq1.4\times10^{12}~\mathrm{GeV}$, respectively.
The individual contributions from the different freeze-in production channels, as well as their sum, are shown separately.

As seen in Fig.~\ref{fig:Yevolve}, for the larger coupling $g=10^{-2}$ (left panel), the process $N_i N_i\to JJ$ dominates the Majoron production, since its cross section scales as $g^4$. On the other hand, for the smaller coupling $g=10^{-4}$ (right panel), the channels $N_i H\to JL_\alpha$, $N_i L_\alpha\to JH$, and $HL_\alpha \to N_iJ$, whose cross sections scale as $g^2|y_{\alpha i}|^2$, provide the dominant contribution.
Among the latter three, the contribution from $HL_\alpha\to N_iJ$ starts to grow earlier than those from $N_iH\to JL_\alpha$ and $N_iL_\alpha\to JH$, since the latter processes require an RHN in the initial state. At late times, these three channels give contributions of the same order, and $N_iH\to JL_\alpha$ becomes somewhat larger.

\begin{figure}
    \centering
    \includegraphics[width=0.7\linewidth]{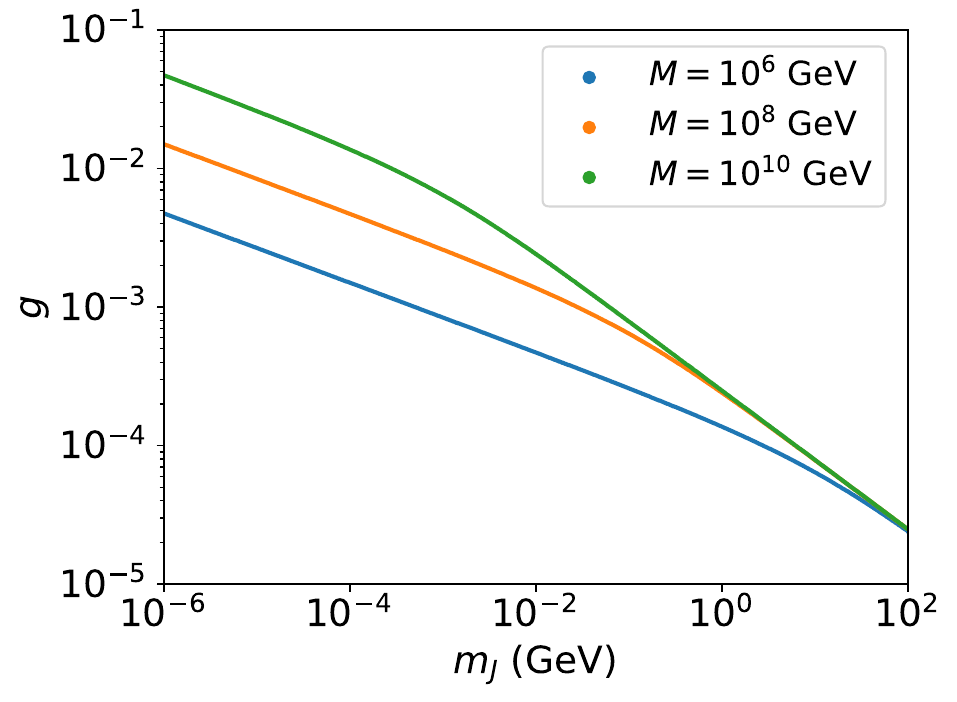}
    \caption{Values of $g$ required to reproduce the observed DM abundance through freeze-in production as a function of the Majoron mass $m_J$, for $M_1=M_2\equiv M$ and $z=0$. The blue, orange, and green curves correspond to $M=10^6$, $10^8$, and $10^{10}\,\mathrm{GeV}$, respectively.}
    \label{fig:mj-g_z=0}
\end{figure}

Fig.~\ref{fig:mj-g_z=0} shows the values of $g$ required for freeze-in (FI) production to reproduce the observed DM abundance, $\Omega_J^{\rm (FI)}h^2 = \Omega_{\rm DM}h^2\simeq 0.12$, as a function of the Majoron mass $m_J$. The different curves correspond to $M=10^6$, $10^8$, and $10^{10}\,\mathrm{GeV}$.

As shown in Fig.~\ref{fig:mj-g_z=0}, for smaller $m_J$, a larger value of $g$ is required to reproduce the observed DM abundance. 
In this regime, the required coupling scales approximately as $g\propto m_J^{-1/4} M^{1/4}$. 
This can be understood from the fact that the production is dominated by the process $N_iN_i\to JJ$, whose production cross section scales as $\sigma\sim g^4/M_i^2$. (See Eq.~\eqref{eq:NN2JJ}.) 
Estimating the Majoron yield at $T\sim M$, one finds $Y_J\sim \sigma n_{N_i}^2 H^{-1}/s \sim g^4/M$, and hence $\Omega_J^{\rm (FI)}h^2\propto m_J Y_J\propto m_J g^4/M$. 
In this regime, we numerically find
\begin{align}
\Omega_J^{\rm (FI)}h^2
&\simeq 0.23
\left(\frac{m_J}{10^{-5}~{\rm GeV}}\right)
\left(\frac{M}{10^8~{\rm GeV}}\right)^{-1}
\left(\frac{g}{10^{-2}}\right)^4
\notag\\
&\simeq 0.94
\left(\frac{m_J}{10^{-5}~{\rm GeV}}\right)
\left(\frac{M}{10^8~{\rm GeV}}\right)^3
\left(\frac{f}{10^{10}~{\rm GeV}}\right)^{-4}
\quad \text{for}\quad
g \gtrsim 10^{-3}\left(\frac{
M}{10^8~{\rm GeV}}\right)^{1/2}.
\label{eq:OmegaFI_large_g}
\end{align}
On the other hand, for larger $m_J$, the required coupling scales approximately as $g\propto m_J^{-1/2}$, and is nearly independent of $M$.
In this regime, the dominant contribution comes from the channels $HL_\alpha\to N_iJ$, $N_iH\to JL_\alpha$, and $N_iL_\alpha\to JH$, 
whose production cross section scales as $\sigma\sim g^2 |y_{\alpha i}|^2/M_i^2$. (See Eqs.~\eqref{eq:NH2JL}--\eqref{eq:HL2NJ}.) Using $|y_{\alpha i}|^2\propto M$, 
one then finds $Y_J\sim \sigma (n_{H/L_\alpha}^{\rm eq})^2 H^{-1}/s\sim g^2$ at $T\sim M$, and hence $\Omega_J^{\rm (FI)}h^2\propto m_J g^2$.
In this regime, we numerically find
\begin{align}
\Omega_J^{\rm (FI)}h^2
&\simeq 0.19
\left(\frac{m_J}{10~{\rm GeV}}\right)
\left(\frac{g}{10^{-4}}\right)^2
\notag\\
&\simeq 0.38
\left(\frac{m_J}{10~{\rm GeV}}\right)
\left(\frac{M}{10^8~{\rm GeV}}\right)^2
\left(\frac{f}{10^{12}~{\rm GeV}}\right)^{-2}
\quad \text{for}\quad
g \lesssim 10^{-3}\left(\frac{
M}{10^8~{\rm GeV}}\right)^{1/2}.
\label{eq:OmegaFI_small_g}
\end{align}
The transition between Eqs.~\eqref{eq:OmegaFI_large_g} and \eqref{eq:OmegaFI_small_g} occurs at 
$g\simeq \mathcal{O}(10^{-3})(M/10^8~{\rm GeV})^{1/2}$
and 
$m_J\simeq\mathcal{O}(10~{\rm MeV})$\\
$(M/10^8~{\rm GeV})^{-1}$.

We close this section with two remarks. First, we note that Majoron DM via freeze-in production is disfavored by structure formation constraints for $m_J \lesssim 3~\mathrm{keV}$~\cite{Villasenor:2022aiy}, though not shown explicitly in Fig.~\ref{fig:mj-g_z=0}.
Second, we comment on the self-consistency of the freeze-in approximation and the existence of an appropriate coupling constant.
In the benchmark examples shown in Figs.~\ref{fig:Yevolve} and \ref{fig:mj-g_z=0}, the Majoron yield remains below the relativistic equilibrium value $Y_J^{\rm eq}\simeq 2.6\times10^{-3}$, so neglecting the inverse processes is reasonable. 
We can also show that, for $m_J\gtrsim \SI{170}{\eV}$, there always exists an appropriate value of $g$ that reproduces the observed DM abundance via freeze-in.
If the coupling were large enough for the Majoron to reach thermal equilibrium, it would 
follow the relativistic freeze-out (FO)
analogously to
SM neutrinos, giving 
$\Omega^{(\mathrm{FO})}_J h^2\simeq 0.12\ (m_J/\SI{170}{\eV})$. 
This exceeds the observed DM abundance $\Omega_{\mathrm{DM}} h^2\simeq 0.12$ for 
$m_J\gtrsim \SI{170}{\eV}$.
On the other hand, the freeze-in yield $\Omega_J^{\mathrm{(FI)}}$ is a monotonically increasing function of $g$, interpolating between $\Omega_J^{\mathrm{(FI)}}\to 0$ as $g\to 0$ and 
$\Omega_J^{\mathrm{(FI)}}\to \Omega_J^{\mathrm{(FO)}}$ as the coupling approaches the equilibration threshold. Therefore, there always exists an appropriate value of $g$
such that $\Omega_J^{\mathrm{(FI)}}=\Omega_{\mathrm{DM}}$, as long as $m_J\gtrsim \SI{170}{\eV}$.

%%%%%%%%%
\subsection{Majoron dark matter via misalignment mechanism}
\label{subsec:Majoron_DM_misalignment}

In addition to the freeze-in production, Majoron DM can also be produced via the misalignment mechanism. 
If the Majoron field is initially displaced from the minimum of its potential, it starts coherent oscillations when the Hubble parameter becomes comparable to the Majoron mass, $3H \simeq m_J$.
At the onset of oscillation, the Majoron energy density is estimated as
\begin{align}
\rho_J^{\rm (osc)} \simeq \frac12 m_J^2 f^2 \theta_i^2 ,
\end{align}
where $\theta_i$ is the initial misalignment angle during inflation. It is expected to be of $\mathcal{O}(1)$ without fine-tuning if the initial angle follows a uniform prior.
In the following, we assume that the onset of the Majoron oscillation takes place during the radiation-dominated era, after reheating is completed. This corresponds to $3H(T_{\rm R}) \gtrsim m_J$, or equivalently
\begin{align}
T_{\rm R}\gtrsim T_{\rm osc}&=
\left(\frac{\pi^2 g_*}{90}\right)^{-1/4}
\sqrt{\frac{m_J M_{\rm Pl}}{3}}
\notag\\
&\simeq
5\times 10^7~{\rm GeV}
\left(\frac{m_J}{0.01~{\rm GeV}}\right)^{1/2},
\end{align}
where $T_\mathrm{osc}$ is the temperature at which the Majoron starts to oscillate.
The ratio of the Majoron energy density to the entropy density, which is conserved after the onset of oscillation, is given by
\begin{align}
\left. \frac{\rho_J}{s}  \right\vert_{\rm osc} 
& = 
\frac{\rho_{\rm rad}}{s} 
\left. \frac{\rho_J}{\rho_{\rm rad}}  \right\vert_{\rm osc}  
\notag\\
&\simeq \frac{3}{4} T_{\rm osc}  
\frac{ (1/2) m_J^2 f^2 \theta_i^2}{3M_{\rm Pl}^{2} (m_J/3)^2}.
\end{align}
This leads to the present abundance
\footnote{Eq.~\eqref{eq:Omega_misalignment} holds for the small-angle regime, and anharmonic corrections enhance the abundance for $\theta_i\simeq\pi$.}
\begin{align}
\Omega_J^{\mathrm{(mis)}} h^2\simeq 0.12\; \theta_i^2 
\left(\frac{f}{4\times 10^{10}~\mathrm{GeV}}\right)^2
\left(\frac{m_J}{10~\mathrm{keV}}\right)^{1/2}.
\label{eq:Omega_misalignment}
\end{align}

%%%%%%%
\subsection{Current experimental constraints and future prospects}
\label{subsec:constraints_and_prospects}

In this subsection, we combine the freeze-in and misalignment contributions;
the total Majoron DM abundance is given by their sum,
$\Omega_J^{(\mathrm{FI})} h^2+\Omega_J^{(\mathrm{mis})} h^2$.
We discuss the parameter region consistent with the observed DM abundance, together with current experimental constraints and future prospects. 
We mainly discuss the case $M_1=M_2=M$ and $z=0$, while more general cases are discussed in Appendix~\ref{sec:appendix_general_case} and the next section.

\begin{figure}[t]
    \centering
    \includegraphics[width=0.45\linewidth]{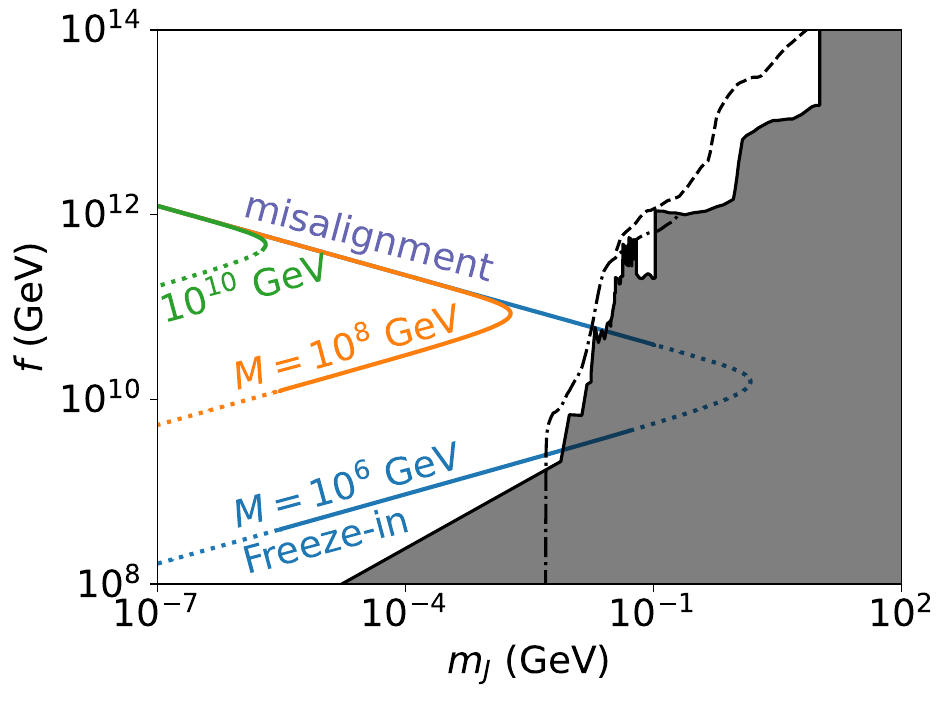}
    \includegraphics[width=0.45\linewidth]{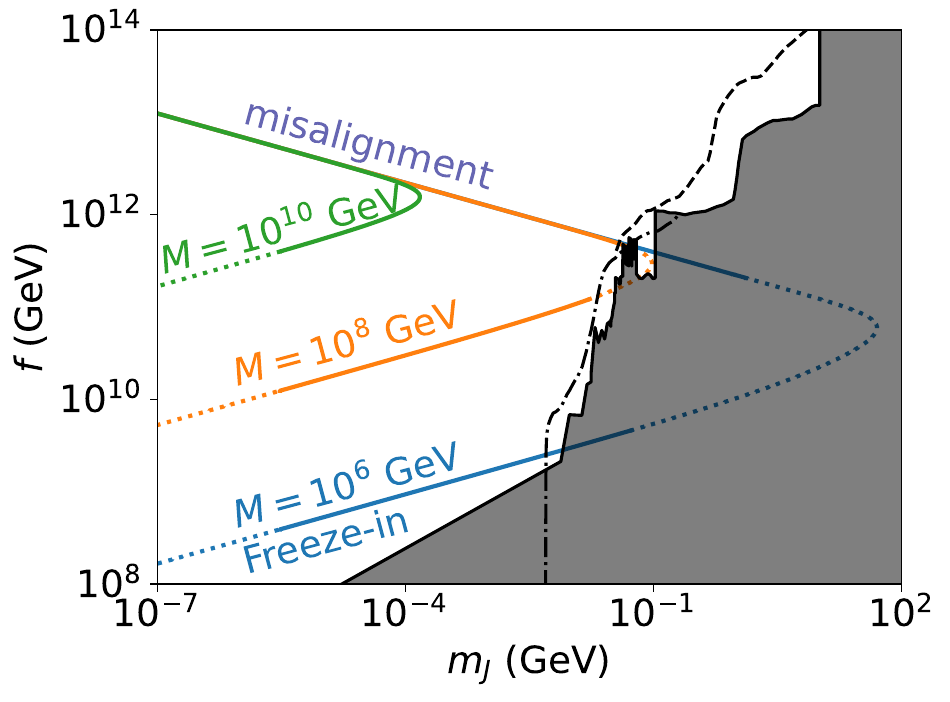}
    \caption{Parameter space in the $(m_J,f)$ plane for $\theta_i=0.1$ (left) and $\theta_i=0.01$ (right), with $M_1=M_2=M$ and $z=0$.
The colored contours show the parameter regions reproducing the observed DM abundance,
    $\Omega_J^{\rm (FI)}h^2 + \Omega_J^{\rm (mis)}h^2 = \Omega_{\rm DM}h^2\simeq 0.12$ for
$M=\SI{e6}{\GeV}$ (blue), $\SI{e8}{\GeV}$ (orange), and $\SI{e10}{\GeV}$ (green).
For each $M$, the contour has two branches: the upper branch is dominated by misalignment production and the lower by freeze-in production.
The dotted portions indicate excluded regions: the lower (freeze-in) branches at $m_J\lesssim 3~\mathrm{keV}$ are excluded by structure formation constraints~\cite{Villasenor:2022aiy}, and those at larger $m_J$ are excluded by the constraints from cosmic-ray and gamma-ray observations (see Fig.~\ref{fig:mj-K_z0}).
The gray shaded region is excluded by current neutrino observations and the cosmological constraint on the DM lifetime from CMB+BAO~\cite{Akita:2023qiz}.
The projected sensitivity in Hyper-Kamiokande (HK)~\cite{Bell:2020rkw} and JUNO~\cite{Akita:2022lit} are also shown as dashed black lines and dot-dashed black lines, respectively~\cite{Akita:2023qiz}.}
    \label{fig:mJ-f}
\end{figure}

Fig.~\ref{fig:mJ-f} shows the resulting parameter space in the $(m_J, f)$ plane for the initial misalignment angle $\theta_i=0.1$ (left) and $\theta_i=0.01$ (right). 
The colored contours indicate the parameter regions in which the total Majoron abundance reproduces the observed DM abundance, 
$\Omega_J^{\rm (FI)}h^2 + \Omega_J^{\rm (mis)}h^2 = \Omega_{\rm DM}h^2\simeq 0.12$. 
The blue, orange, and green curves correspond to $M=\SI{e6}{\GeV}$, $\SI{e8}{\GeV}$, and $\SI{e10}{\GeV}$, respectively.
For each value of $M$, the contour consists of two branches: the upper branch is dominated by misalignment production, while the lower branch is dominated by freeze-in production. 
The dotted portions of the lower (freeze-in) branches at $m_J\lesssim 3~\mathrm{keV}$ 
are excluded by structure formation constraints~\cite{Villasenor:2022aiy}.\footnote{For $\theta_i=0.1$ and 
$M=10^{10}~\rm{GeV}$, the dotted line corresponds to 
$\Omega_J^{\rm{(FI)}}>\Omega_J^{\rm{(mis)}}$.}
The dotted portions at larger $m_J$, the gray shaded regions, and the black dashed lines indicate observational constraints and future sensitivities, which are discussed below.

As shown in Fig.~\ref{fig:mJ-f}, the abundance contours exhibit a characteristic behavior reflecting the interplay between the freeze-in and misalignment contributions. 
For fixed $M$, the freeze-in contribution increases as $f$ decreases, as can be seen from Eqs.~\eqref{eq:OmegaFI_large_g} and \eqref{eq:OmegaFI_small_g}, while the misalignment contribution grows with $f$, as shown in Eq.~\eqref{eq:Omega_misalignment}.
As a result, the condition $\Omega_J^{\rm (FI)}h^2+\Omega_J^{\rm (mis)}h^2=\Omega_{\rm DM}h^2$ is satisfied only in a finite region of the $(m_J,f)$ plane.
We also note that, except in the large $m_J$ region, the freeze-in contribution is dominated by the $N_iN_i \to JJ$ process, as discussed in Sec.~\ref{subsec:Freeze-in}.

We now turn to the experimental constraints shown in Fig.~\ref{fig:mJ-f}.
Majoron decays into neutrinos at the tree level, mediated by the mixing between active and RHNs.
Furthermore, it decays into leptons and quarks at the one-loop level, and into photons at the two-loop level.
The decay rate for each process is as follows\cite{Pilaftsis:1993af,Garcia-Cely:2017oco, Heeck:2019guh}: 
\begin{align}
    \Gamma_{J\rightarrow\nu\nu}
    &\simeq\frac{m_J}{16\pi f^2}\sum_{i=1}^3m_{\nu_i}^2,\label{eq:decay_nu}\\
    \Gamma_{J\rightarrow q\bar{q}}
    &\simeq\frac{3m_J}{8\pi}\left|\frac{m_q}{8\pi^2v}T^q_3\tr K\right|^2,\label{eq:decay_q}\\
    \Gamma_{J\rightarrow l\bar{l}}
    &\simeq\frac{m_J}{8\pi}\left|\frac{m_l}{8\pi^2v}(T_3^l\tr K+K_{ll})\right|^2\label{eq:decay_l}\\
    \Gamma_{J\rightarrow \gamma\gamma}
    &\simeq\frac{\alpha^2}{4096\pi^7}\frac{m_J^3}{v^2}|K'|^2,\label{eq:decay_photon}
\end{align}
with
\begin{align}
    K&=\frac{m_Dm_D^\dagger}{vf},
    \label{eq:K}
    \\
    K'&=\tr K\sum_\psi N_c^\psi T_3^\psi Q_\psi ^2h\left(\frac{m_J^2}{4m_\psi^2}\right)+\sum_lK_{ll}h\left(\frac{m_J^2}{4m_l^2}\right), 
    \label{eq:K'}
\end{align}
where $\psi$ runs over SM fermions,\footnote{For $m_J \lesssim \Lambda_{\mathrm{QCD}}$, light-quark loops should be replaced by hadronic contributions. For simplicity, we neglect these contributions and keep only the lepton loops for $m_J<1~\mathrm{GeV}$.
See also Ref.~\cite{Bauer:2017ris}.
}
with $u$ denoting up-type quarks ($u,c,t$), $d$ down-type quarks ($d,s,b$), and $l$ charged leptons ($e,\mu,\tau$), $T_3^u=-T_3^{d,l}=1/2$ is the isospin, $N_c^{u,d}=3$ and $N_c^l=1$ are the number of colors, $Q_\psi$ is the electric charge, $\alpha$ is the fine-structure constant, and
\begin{align}
    h(x)&=-\frac{1}{4x}\left(\log(1-2x+2\sqrt{x(x-1)})\right)^2-1.
    \label{eq:h(x)}
\end{align}
In Eqs.~(\ref{eq:decay_nu})-(\ref{eq:decay_photon}), we have assumed that the RHNs are much heavier than the other relevant particles.

\begin{figure}[!t]
    \centering
    \includegraphics[width=0.45\linewidth]{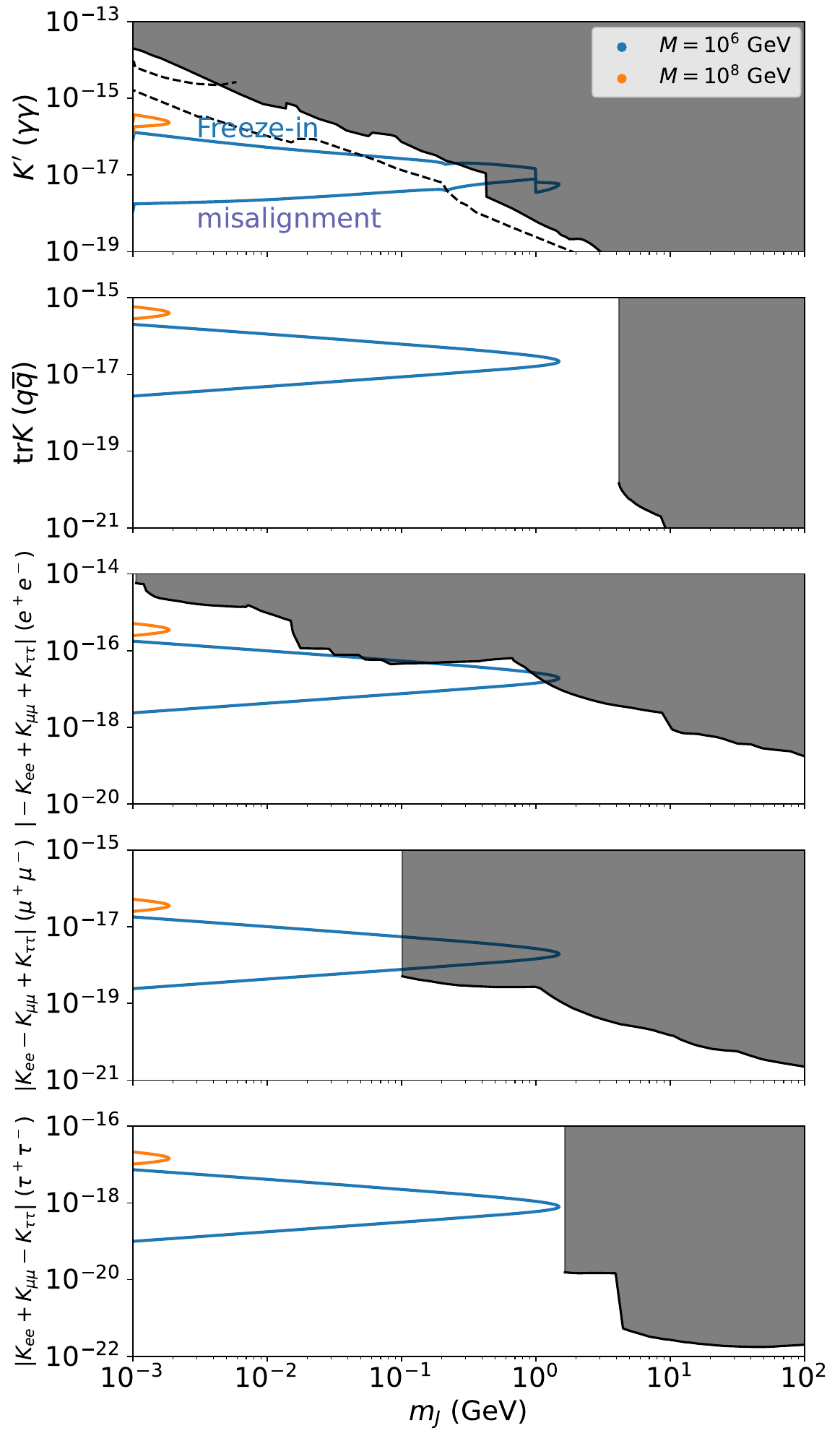}
    \includegraphics[width=0.45\linewidth]{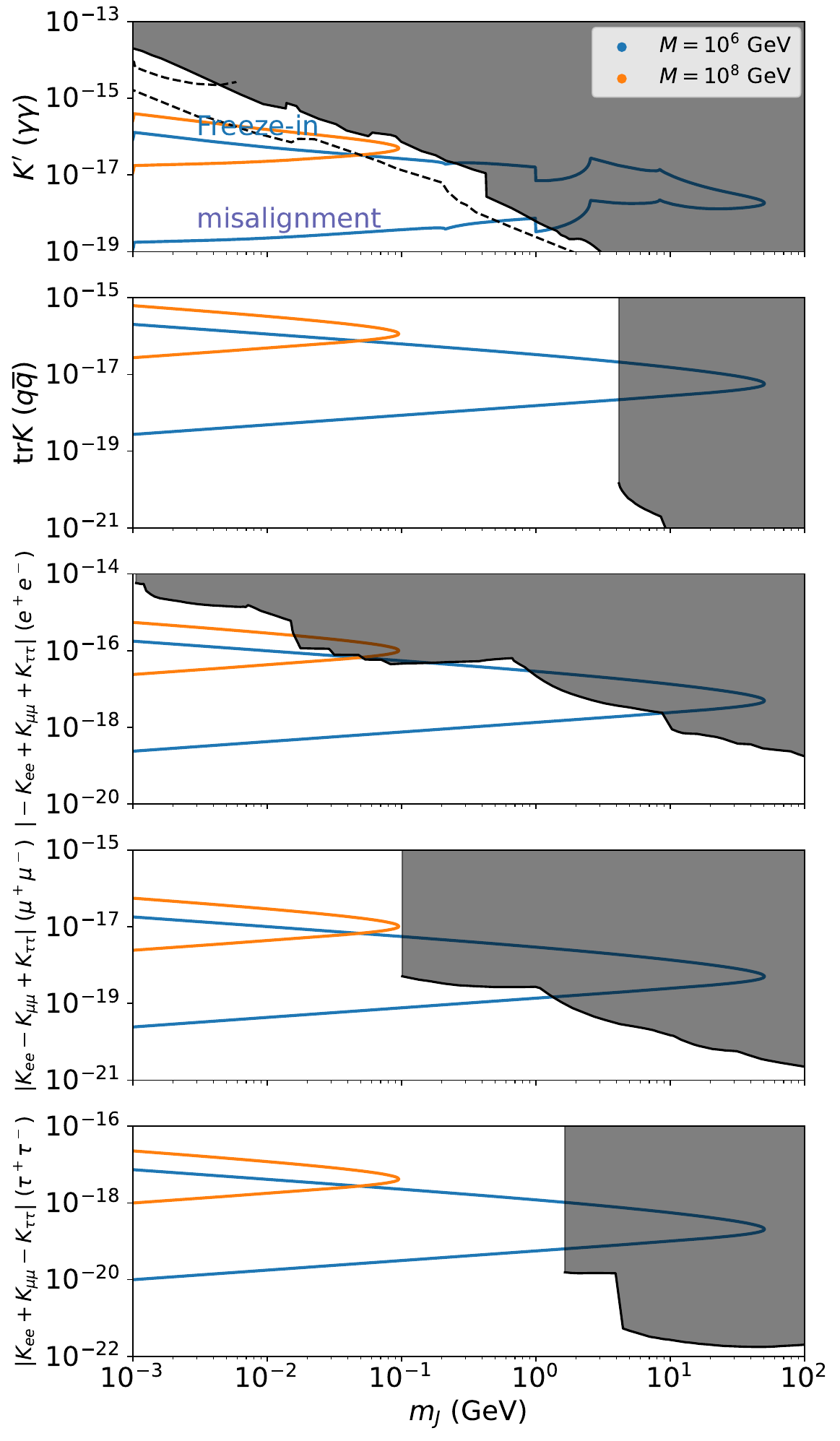}
    \caption{Values of various combinations of $K$ required to reproduce the observed DM abundance as a function of $m_J$, with $M_1=M_2=M$ and $z=0$, for $\theta_i=0.1$ (left) and $\theta_i=0.01$ (right). 
The blue and orange curves correspond to $M=\SI{e6}{\GeV}$ and $\SI{e8}{\GeV}$, respectively, following the same convention as in Fig.~\ref{fig:mJ-f}; the green curve ($M=\SI{e10}{\GeV}$) lies outside the displayed mass range. 
For each curve, the upper branch corresponds to freeze-in-dominated production and the lower to misalignment-dominated production.
The gray shaded regions indicate excluded parameter regions; 
constraints on $J\rightarrow\gamma\gamma$ are from Fermi-LAT~\cite{Fermi-LAT:2015kyq}, INTEGRAL/SPI~\cite{Fischer:2022pse} with NFW DM profile,  COMPTEL/EGRET~\cite{Essig:2013goa},
and those on $J\rightarrow q\bar{q}$ and $J\rightarrow l\bar{l}$ 
are from \cite{Akita:2023qiz}. 
The portions of the colored contours overlapping with the gray shaded regions correspond to the dotted portions at larger $m_J$ in Fig.~\ref{fig:mJ-f}.
The dashed line in the top panels shows the projected sensitivities of future gamma-ray observations: COSI~\cite{Caputo:2022dkz,Tomsick:2023aue} (upper) and other proposed missions summarized in Ref.~\cite{ODonnell:2024aaw} (lower).
}
    \label{fig:mj-K_z0}
\end{figure}

In Fig.~\ref{fig:mJ-f}, the gray shaded region is excluded by current 
neutrino observations and the cosmological constraint on the DM lifetime 
from CMB+BAO~\cite{Akita:2023qiz}. 
These constraints bound the parameter $f$ through the decay rate 
$\Gamma_{J\to\nu\nu}$ in Eq.~\eqref{eq:decay_nu}. 
The projected sensitivities of Hyper-Kamiokande (HK) \cite{Bell:2020rkw} and JUNO \cite{Akita:2022lit} are also 
shown as dashed black lines and dot-dashed black lines, respectively~\cite{Akita:2023qiz}, which can probe part of 
the viable parameter space.
See also Ref.~\cite{Arguelles:2022nbl} for the constraints and future sensitivities on the lifetime of heavier DM decaying to neutrinos.

On the other hand, the decay rates into quarks, leptons, and photons in Eqs.~\eqref{eq:decay_q}--\eqref{eq:decay_photon} constrain the parameter $K$ rather than $f$ directly. 
Fig.~\ref{fig:mj-K_z0} shows the values of the relevant combinations of $K$ required to reproduce the observed DM abundance as a function of $m_J$, with $M_1=M_2=M$ and $z=0$, for $\theta_i=0.1$ (left) and $\theta_i=0.01$ (right), following the same color convention as in Fig.~\ref{fig:mJ-f}.
Note that the green curve ($M=10^{10}~\mathrm{GeV}$) is not shown since it lies outside the mass range displayed.
For each curve, the upper branch corresponds to freeze-in-dominated production and the lower to misalignment-dominated production.
The gray shaded regions show current observational constraints; 
constraints on $J\rightarrow\gamma\gamma$ are from Fermi-LAT~\cite{Fermi-LAT:2015kyq}, INTEGRAL/SPI~\cite{Fischer:2022pse} with NFW DM profile,  COMPTEL/EGRET~\cite{Essig:2013goa},
and those on $J\rightarrow q\bar{q}$ and $J\rightarrow l\bar{l}$ 
are from \cite{Akita:2023qiz}.
The parameter regions where the colored contours enter the gray shaded region correspond to the dotted portions at larger $m_J$ in Fig.~\ref{fig:mJ-f}.
The dashed line in the top panels shows the projected sensitivities of future gamma-ray observations:
COSI~\cite{Caputo:2022dkz,Tomsick:2023aue} (upper) and other proposed missions summarized in Ref.~\cite{ODonnell:2024aaw} (lower). 
One can see that a part of the viable parameter space can be probed by future $\gamma$-ray observations.

For the misalignment-dominated case, the Majoron acquires an isocurvature perturbation, which is constrained by Planck observations\cite{Planck:2018jri}
\footnote{
As in Eq.~(\ref{eq:Omega_misalignment}), Eq.~(\ref{eq:iso}) receives anharmonic corrections for $\theta_i\simeq\pi$.
}
\begin{align}
    \beta_{\rm{iso}}&=\left(1+\frac{\pi^2A_sf^2\theta_i^2}{H_{\rm{inf}}^2}\right)^{-1}
    \simeq\frac{H_{\rm{inf}}^2}{\pi^2A_sf^2\theta_i^2}<0.038
    \label{eq:iso}
\end{align}
where $A_s=2.1\times10^{-9}$\cite{Planck:2018vyg}. From this, we obtain an upper bound on $H_{\rm{inf}}$, 
\begin{align}
    H_{\rm{inf}}&\lesssim3\times10^6\left(\frac{f}{10^{12}~\rm{GeV}}\right)\left(\frac{\theta_i}{0.1}\right)~\rm{GeV}\notag\\
    &\simeq1.1\times10^7\left(\frac{\Omega_J h^2}{0.12}\right)^{1/2}\left(\frac{m_J}{1~\rm{eV}}\right)^{-1/4}~\rm{GeV}.
    \label{eq:H_iso}
\end{align}

For instantaneous reheating, this leads to
\begin{align}
    T_{R}&=\left(\frac{90}{\pi^2g_*}\right)^{1/4}\sqrt{H_{\rm{inf}}M_{\rm{Pl}}}\notag\\
    &\lesssim3\times10^{12}\left(\frac{m_J}{1~\rm{eV}}\right)^{-1/8}~\rm{GeV}.
    \label{eq:T_iso}
\end{align}
Since our analysis assumes $T_R=10M$, the reheating temperature is sufficiently low that the isocurvature constraint derived here is irrelevant in both panels of Fig.~\ref{fig:mJ-f}. If the reheating temperature is increased toward $T_R\sim f$, the isocurvature constraint would exclude the small-$m_J$ end of the misalignment branch in the right panel of Fig.~\ref{fig:mJ-f}, where $f\gtrsim 10^{13}~\mathrm{GeV}$.

As can be seen from Figs.~\ref{fig:mJ-f} and \ref{fig:mj-K_z0}, 
the Majoron mass is bounded from above by the observational constraints. 
Without severe fine-tuning of the initial misalignment angle, we find $m_J \lesssim \mathcal{O}(10)~\mathrm{MeV}$ depending on $\theta_i$ and $M$.
Note that the dotted portions at larger $m_J$ in Fig.~\ref{fig:mJ-f} are largely covered by the neutrino constraints (gray shaded region), so the latter provide the dominant upper bound on $m_J$.
Furthermore, for the freeze-in dominated case, the requirement of 
reproducing the observed DM abundance also implies an upper bound on 
the RHN mass scale $M$, as can be seen from Fig.~\ref{fig:mJ-f}. 
This will be further discussed in Sec.~\ref{sec:Majoron_DM_and_LG}.

%%%%%
\section{Majoron dark matter and leptogenesis}
\label{sec:Majoron_DM_and_LG}

In this section, we discuss the implications of Majoron DM for thermal leptogenesis. 
It is known that successful thermal leptogenesis requires sufficiently large RHN masses; in the standard scenario with hierarchical RHN masses, 
$M\gtrsim 10^9~\mathrm{GeV}$~\cite{Davidson:2002qv}, and more stringent in the minimal framework with two RHNs~\cite{Chankowski:2003rr, Guo:2006qa, Blanchet:2008pw}. (See also the discussion below.)

A qualitative picture can already be obtained from Fig.~\ref{fig:mJ-f}. Without fine-tuning of $\theta_i$, the parameter region compatible with both the observed DM abundance and large $M$ tends to favor small Majoron masses and misalignment-dominated production. The freeze-in branch requires smaller values of $M$, leaving limited room for compatibility with thermal leptogenesis. 
However, since Fig.~\ref{fig:mJ-f} assumes $M_1=M_2$ and $z=0$, for which leptogenesis does not work, a more careful analysis is needed.

To this end, we consider the following three benchmark points:
\begin{align}
(M_1, M_2, a, b) &= (10^{12}~\mathrm{GeV},\ 10^{13}~\mathrm{GeV},\ 1.6,\ \pi/4),
\notag\\
(M_1, M_2, a, b) &= (10^{11}~\mathrm{GeV},\ 10^{12}~\mathrm{GeV},\ 1.9,\ \pi/8),
\notag\\
(M_1, M_2, a, b) &= (2.7\times10^{10}~\mathrm{GeV},\ 2.7\times10^{11}~\mathrm{GeV},\ 2.8,\ 0.28),
\label{eq:LG_benchmark}
\end{align}
for which we have explicitly verified that the observed baryon asymmetry is generated via leptogenesis using the public code \texttt{ULYSSES}~\cite{Granelli:2020pim,Granelli:2023vcm}.
For simplicity, we set $M_2 = 10 M_1$ and $T_{\rm{R}} = M_2$. The latter choice ensures that the $\mathrm{U(1)_{B-L}}$
symmetry is not restored after reheating.\footnote{This follows from $g_2 f/\sqrt{2} \lesssim f$ since $g_2 \lesssim 1$.}
The neutrino oscillation parameters are chosen as discussed in Sec.~\ref{sec:setup}, and the Majorana phase is set to $\alpha=0$. 
The third benchmark point corresponds to the lowest RHN mass for which we found successful leptogenesis under these assumptions.\footnote{$M_1\simeq 2.7\times 10^{10}~\mathrm{GeV}$ is somewhat below the bound reported in Refs.~\cite{Chankowski:2003rr, Guo:2006qa, Blanchet:2008pw} for the minimal two-RHN framework, which might be due to the flavor effects.
We also note that we have not performed an exhaustive parameter scan.}

\begin{figure}[t]
    \centering
    \includegraphics[width=0.7\linewidth]{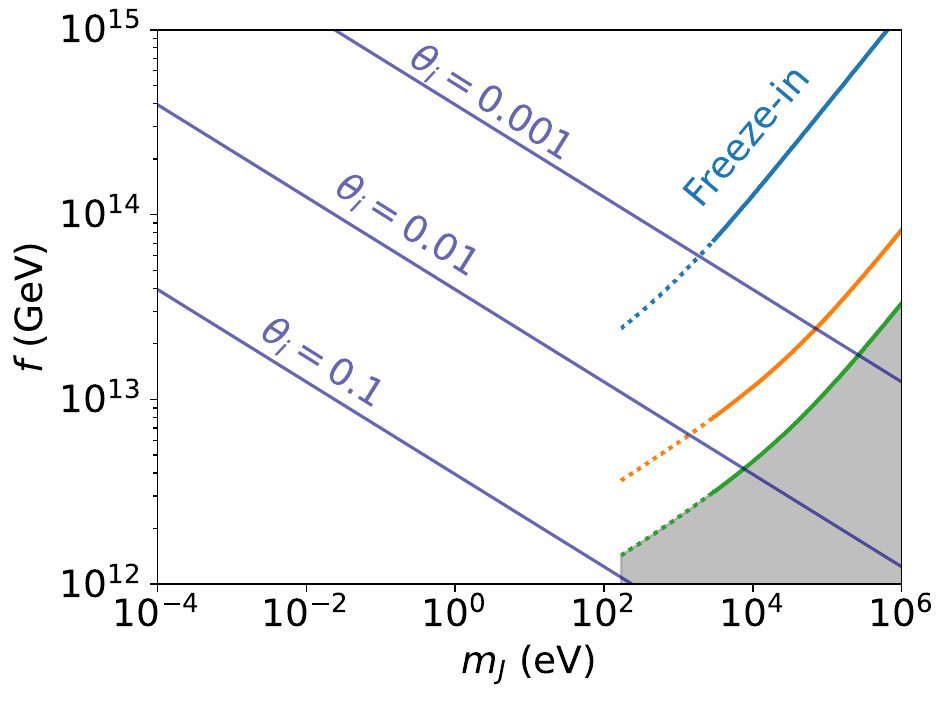}
    \caption{Values of $f$ required to reproduce the observed DM abundance via the misalignment mechanism (navy curves, for $\theta_i = 0.1$, $0.01$, and $0.001$ from bottom to top) and via freeze-in production (colored curves) as a function of $m_J$, with $M_2 = 10M_1$ and $T_{\rm{R}} = M_2$. 
    The blue, orange, and green curves correspond to the three benchmark points with $M_1 = 10^{12}$, $10^{11}$, and $2.7\times10^{10}~\mathrm{GeV}$, respectively, for which successful leptogenesis has been verified (see text for details). 
    The dotted portions indicate the region excluded by structure formation constraints~\cite{Villasenor:2022aiy}. 
    The gray shaded region indicates where freeze-in overproduces DM even for the lowest leptogenesis-compatible RHN mass, and is therefore disfavored by leptogenesis.}
    \label{fig:mj-f_LG}
\end{figure}

Fig.~\ref{fig:mj-f_LG} shows the values of $f$ required to reproduce the observed DM abundance via the misalignment mechanism (navy curves, for $\theta_i = 0.1$, $0.01$, and $0.001$ from bottom to top) and via freeze-in production (colored curves) as a function of $m_J$. The blue, orange, and green curves correspond to the three benchmark points in Eq.~\eqref{eq:LG_benchmark}. The freeze-in curves start at $m_J \simeq \SI{170}{\eV}$, below which the Majoron thermalizes and the freeze-in mechanism cannot reproduce the observed DM abundance (see Sec.~\ref{subsec:Freeze-in}). The dotted portions indicate the region excluded by structure formation constraints~\cite{Villasenor:2022aiy}.
\footnote{
We note that thermally produced light Majorons behave as dark radiation, which contribute to the effective number of neutrino species, $\Delta N_{\rm eff}$. The BBN and CMB constraints on the dark radiation are $\Delta N_{\rm eff}\lesssim 0.107$ \cite{Yeh:2026pil,Goldstein:2026iuu} at the freeze-out epoch of proton-neutron ratio and $\Delta N_{\rm eff}\lesssim 0.3$ \cite{Planck:2018vyg} at the epoch of matter-radiation equality, respectively. The maximum contribution of the Majoron to $\Delta N_{\rm{eff}}$ is achieved in the relativistic freeze-out regime, giving $\Delta N_{\rm{eff}}=(4/7)(10.75/106.75)^{4/3}\simeq0.027$.
This is much lower than the observational bound.
}
The gray shaded region indicates where the freeze-in mechanism overproduces DM even for the lowest leptogenesis-compatible RHN mass, and is therefore disfavored by leptogenesis. On the other hand, the region above the navy lines indicates where the misalignment mechanism overproduces DM.

From the figure, for Majoron DM produced predominantly via freeze-in, 
successful thermal leptogenesis requires $\theta_i \lesssim \mathcal{O}(0.01)$ 
to avoid overproduction from the misalignment mechanism. 
This implies that the simultaneous realization of freeze-in Majoron DM 
and successful thermal leptogenesis requires a mild fine-tuning 
of the initial misalignment angle.
Note that, in this parameter region, the freeze-in production is dominated by the $N_iN_i\to JJ$ process (see Sec.~\ref{subsec:Freeze-in}), which is independent of the Casas-Ibarra parameters. The conclusion above is therefore 
robust against the specific choice of benchmark points.

On the other hand, if Majoron DM is produced predominantly via misalignment, the viable parameter space is significantly broader, with no lower bound on $m_J$ from freeze-in considerations. 
In particular, for $m_J \lesssim 100~\mathrm{eV}$, the freeze-in contribution is automatically suppressed, and Majoron DM can be produced via the misalignment mechanism without any constraint from freeze-in overproduction.

For the misalignment-produced Majoron DM, isocurvature perturbations impose the upper bound on the reheating temperature as shown in Eq.~(\ref{eq:T_iso}). 
Under the assumption of instantaneous reheating with $T_R=M_2$, the isocurvature bound excludes the combination of misalignment-dominated Majoron DM and the highest-scale benchmark, $M_2=10^{13}~\mathrm{GeV}$, over the mass range shown in Fig.~5. By contrast, the other two benchmarks, with $M_2=10^{12}~\mathrm{GeV}$ and $2.7\times10^{11}~\mathrm{GeV}$, satisfy the isocurvature bound in the misalignment-dominated region, $m_J\lesssim100~\mathrm{eV}$.

For thermal leptogenesis with three RHNs or low-scale leptogenesis with degenerate RHN masses, the parameter space for Majorons compatible with the observed DM abundance may be further opened up. We will leave the detailed analysis for future work.

%%%%
\section{Conclusion}
\label{sec:conclusion}

In this paper, we considered Majoron DM in its minimal setup. Our analysis is based on the Type-I seesaw framework with two RHNs, extended by a SM-singlet complex scalar. We assume that the global $\mathrm{U(1)_{B-L}}$ is explicitly broken by high-energy physics, which generates a Majoron mass and makes it a viable DM candidate. Throughout the present work, we remained agnostic about the origin of the explicit breaking of the global $\mathrm{U(1)_{B-L}}$ and treated the Majoron mass as a free parameter.

In this minimal setup, we studied the production of Majoron DM through both the freeze-in mechanism and the misalignment mechanism, and evaluated the resulting DM abundance. 
We identified the viable parameter regions for Majoron DM, which are summarized in Fig.~\ref{fig:mJ-f}. 
We found that, without severe fine-tuning of the initial misalignment angle (taking $\theta_i\simeq 0.1$ and $0.01$ as representative values), the Majoron mass is constrained to satisfy $m_J\lesssim \mathcal{O}(10)~\si{\MeV}$.
Part of the viable parameter space is within reach of future neutrino experiments such as Hyper-Kamiokande and JUNO, as well as future $\gamma$-ray observations.

We also studied the implications of Majoron DM for thermal leptogenesis in the minimal setup with two RHNs. 
Fig.~\ref{fig:mj-f_LG} shows that, in order to simultaneously achieve freeze-in production of Majoron DM and successful thermal leptogenesis, we need a mild fine-tuning in the initial misalignment angle, $\theta_i\lesssim 0.01$. 
On the other hand, the misalignment mechanism is compatible with thermal leptogenesis with $m_J\lesssim \mathcal{O}(100)~{\rm eV}$ without any fine-tuning.

Our discussion in the present paper is limited to the case with $f > T_{\rm{R}}$, where the global $\mathrm{U(1)_{B-L}}$ symmetry is not thermally restored after reheating. Investigating the case with $f < T_{\rm{R}}$ is left for future work. For leptogenesis, we only considered the thermal scenario with two RHNs. If we consider thermal leptogenesis with three RHNs or low-scale leptogenesis, part of the viable parameter space for Majoron DM consistent with the observed BAU would be reopened.
We will leave the detailed analysis for future work. 
\\
\\
\textit{\textbf{Note added.}}---Shortly after our paper appeared on arXiv, related work \cite{deGiorgi:2026jqn,Batell:2026avi} appeared, discussing a similar scenario.

\section*{Acknowledgments}
\noindent

We thank Subaru Fujisawa for pointing out an error in an equation in the previous version of this work.
This work was supported by JSPS KAKENHI Grant Numbers 24H02244, 24K07041 (KH), 24KJ0060 (KA), and 25KJ0779 (TY).

%%%
\appendix

\section{Majoron Production and Decay with $M_1\neq M_2$ and $z\neq 0$}
\label{sec:appendix_general_case}

In Sec.~\ref{sec:production}, we have studied the Majoron DM parameter
space using the simple benchmark choice $M_1=M_2$ and $z=0$.
In this appendix, we discuss how the production and decay of the Majoron
are affected when these benchmark assumptions are relaxed.
The purpose of this appendix is not to perform a complete scan over the
full parameter space, but to clarify how the main conclusions of
Sec.~\ref{sec:production} are modified by nondegenerate RHN masses and a
nonzero Casas-Ibarra parameter.

It is useful to first recall the parametric dependence of the relevant
processes.  In Sec.~\ref{sec:production}, the freeze-in abundance was
mainly controlled by the process $N_iN_i\to JJ$ (Eq.~\eqref{eq:NN2JJ}), while the misalignment
contribution was determined by $f$, $m_J$, and the initial misalignment
angle (Eq.~\eqref{eq:Omega_misalignment}).  
The strongest decay constraint in most of the parameter space came from the tree-level decay $J\to\nu\nu$, whose rate is fixed by the
light-neutrino masses and $f$ (Eq.~\eqref{eq:decay_nu}).
These production mechanisms and the tree-level decay are independent of
the Casas-Ibarra parameter $z$.
The dependence on $z$ appears mainly through loop-induced decay modes,
which involve the parameter $K$ in Eq.~\eqref{eq:K}, and through
Yukawa-mediated production channels such as $HL_\alpha\to N_iJ$ (Eqs.~\eqref{eq:NH2JL}--\eqref{eq:HL2NJ}).

We first discuss the effect of a nondegenerate RHN mass spectrum,
focusing on the $z$-independent processes.
In the $(m_J, f)$ plane of Fig.~\ref{fig:mJ-f}, the neutrino constraint 
and the misalignment contribution are independent of the RHN masses 
and remain unchanged.
The freeze-in contribution via $N_iN_i\to JJ$ depends on the RHN masses as
$\Omega_J^{\rm (FI)}\sim m_J f^{-4} M^3$, as shown in Eq.~\eqref{eq:OmegaFI_large_g}.
When $M_1\ll M_2\ll T_{\rm{R}}$, both RHNs are thermalized, and the production 
is dominated by the heavier RHN $N_2$; the results of Sec.~\ref{sec:production} apply with $M\to M_2$, up to an overall factor of $1/2$ in the abundance compared with the degenerate benchmark.
In contrast, when $M_1\ll T_{\rm{R}}\ll M_2$, only $N_1$ is thermalized, and the freeze-in production is dominated by $N_1$; the results apply with $M\to M_1$, again up to the same factor.
In both cases, the qualitative features of Fig.~\ref{fig:mJ-f} remain 
largely unchanged, and in particular the upper bound on $m_J$ is not 
significantly affected, up to additional constraints from the loop-induced decay modes discussed below.

We next consider the 
Yukawa-mediated channels $HL_\alpha\to N_iJ$, $N_iH\to JL_\alpha$, 
and $N_iL_\alpha\to JH$.
As shown in Eqs.~\eqref{eq:NH2JL}--\eqref{eq:HL2NJ}, the corresponding production cross sections are proportional to $\sum_\alpha |y_{\alpha i}|^2=(y^\dagger y)_{ii} \propto (m_D^\dagger m_D)_{ii}$. Using Eq.~\eqref{eq:CI-para}, one finds
\begin{align}
(m_D^\dagger m_D)_{11}
&=M_1\left(\hat{m}^\nu_2\frac{\cos2a+\cosh2b}{2}-\hat{m}^\nu_3\frac{\cos2a-\cosh2b}{2}\right),
\\
(m_D^\dagger m_D)_{22}
&=
M_2\left(\hat{m}^\nu_3\frac{\cos2a+\cosh2b}{2}-\hat{m}^\nu_2\frac{\cos2a-\cosh2b}{2}\right),
\end{align}
which can be exponentially enhanced for large $|b|$.
For small $|b|$, these channels are subdominant compared to $N_iN_i\to JJ$, 
and the results of Sec.~\ref{sec:production} remain a good approximation.
As $|b|$ increases, however, the Yukawa-mediated channels become dominant, 
and the freeze-in abundance is enhanced. 
In the $(m_J, f)$ plane of Fig.~\ref{fig:mJ-f}, this shifts the 
freeze-in contour upward, reducing the viable parameter space.

Finally, we turn to the loop-induced decay modes, which depend on the parameter $K$ in Eq.~\eqref{eq:K}.
Using Eq.~\eqref{eq:CI-para}, one finds
\begin{align}
    &vfK\notag\\
    &=m_Dm_D^\dagger\notag\\
    &=U\mqty(0&0&0\\0
    &\hat{m}^\nu_2\left(M_1\frac{\cos2a+\cosh2b}{2}-M_2\frac{\cos2a-\cosh2b}{2}\right)
    &\sqrt{\hat{m}^\nu_2\hat{m}^\nu_3}\left(\frac{M_1-M_2}{2}\sin2a-i\frac{M_1+M_2}{2}\sinh2b\right)\\0
    &\sqrt{\hat{m}^\nu_2\hat{m}^\nu_3}\left(\frac{M_1-M_2}{2}\sin2a+i\frac{M_1+M_2}{2}\sinh2b\right)
    &\hat{m}^\nu_3\left(M_2\frac{\cos2a+\cosh2b}{2}-M_1\frac{\cos2a-\cosh2b}{2}\right))U^\dagger
    \label{eq:K_expr}
\end{align}
From this expression, one sees that $K$ depends on both the RHN mass spectrum and the Casas-Ibarra parameter. 
Regarding the RHN mass hierarchy, when $M_1\ll M_2\ll T_{\rm{R}}$, both the freeze-in production and $K$ are dominated by the heavier RHN $N_2$, so the constraints are not significantly affected by the mass hierarchy.
In contrast, when $M_1\ll T_{\rm{R}}\ll M_2$, the freeze-in production is dominated by $N_1$, while $K$ receives contributions mainly from the heavier RHN $N_2$ and is enhanced by a factor of approximately $M_2/M_1\gg 1$.
The constraints on $K$ therefore tend to become considerably tighter than in the benchmark case of Sec.~\ref{sec:production}.

\begin{figure}[t]
    \centering
    \includegraphics[width=0.45\linewidth]{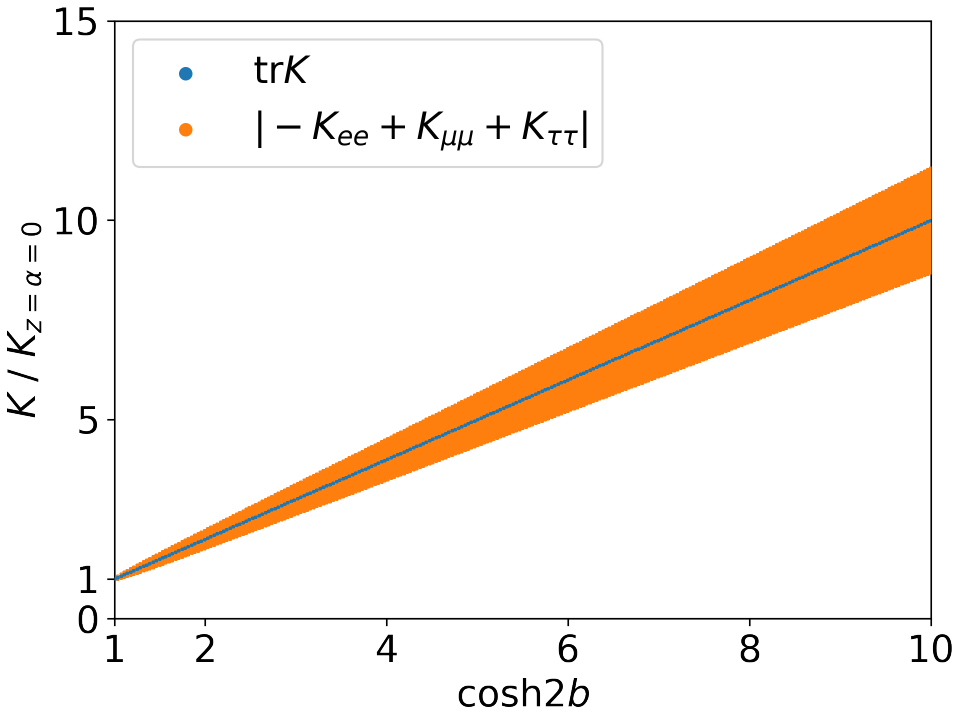}
    \includegraphics[width=0.45\linewidth]{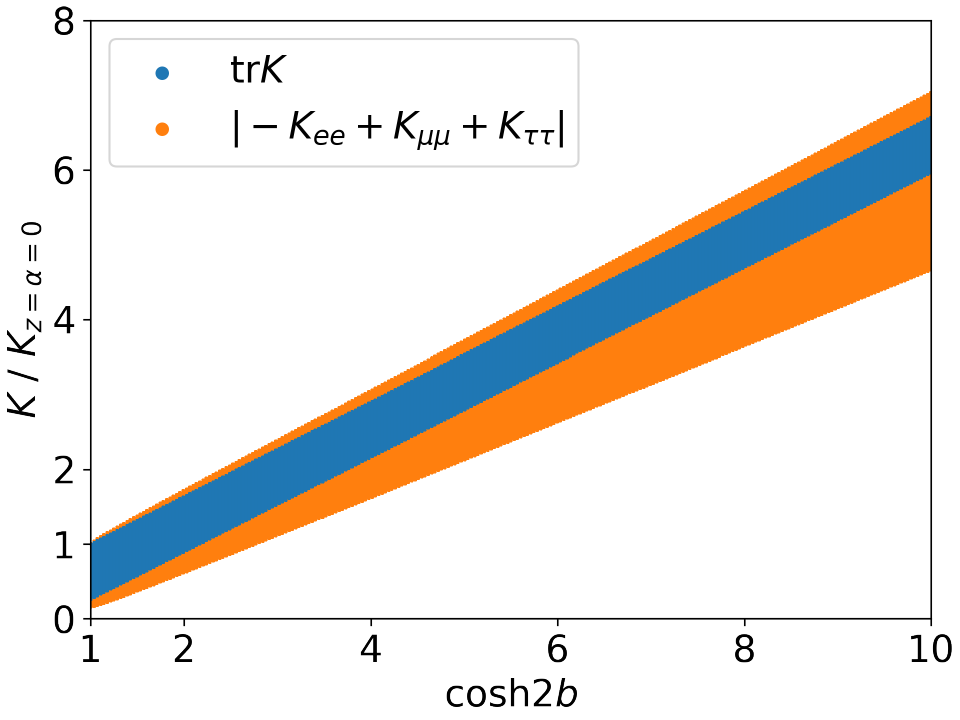}
    \caption{Dependence of $\tr K$ and $|-K_{ee}+K_{\mu\mu}+K_{\tau\tau}|$ 
on the Casas-Ibarra parameter $z=a+ib$ and the Majorana phase $\alpha$, for $M_2/M_1=1$ (left) and $M_2/M_1=10$ (right).
The bands show the range obtained by varying $a\in[0,\pi]$ and $\alpha\in[0,2\pi]$ 
for each value of $b$, normalized to the benchmark case $z=\alpha=0$.
}
\label{fig:trK-Kee}
\end{figure}

We now examine the dependence of $K$ on the Casas-Ibarra parameter $z=a+ib$ and the Majorana phase $\alpha$ in the PMNS matrix $U$.
From Eq.~\eqref{eq:K_expr}, the entries of $K$ scale as $\cosh 2b$ for large $|b|$, so larger values of $|b|$ generically lead to more stringent constraints from the loop-induced decay modes.
However, cancellations among $K_{ee}$, $K_{\mu\mu}$, and $K_{\tau\tau}$ can occur for specific values of $a$ and $\alpha$, weakening these constraints.
Fig.~\ref{fig:trK-Kee} illustrates how $\tr K$ and $|-K_{ee}+K_{\mu\mu}+K_{\tau\tau}|$ (which govern the decays into quarks and electrons) depend on $z=a+ib$ and $\alpha$ for RHN mass hierarchies of $M_2/M_1=1,10$ (left and right).
The bands show the range obtained by varying $a\in[0,\pi]$ and $\alpha\in[0,2\pi]$ for each value of $b$, normalized to the benchmark case $z=\alpha=0$.
The figure highlights the dominant dependence on $b$ alongside the variance introduced by $a$ and $\alpha$.
From Fig.~\ref{fig:trK-Kee}, we find that while $\tr K$ and $|-K_{ee}+K_{\mu\mu}+K_{\tau\tau}|$ are mainly governed by $b$, the effects of $a$ and $\alpha$ only introduce $\mathcal{O}(1)$ corrections.
In contrast, $|K_{ee}-K_{\mu\mu}+K_{\tau\tau}|$ and 
$|K_{ee}+K_{\mu\mu}-K_{\tau\tau}|$, which govern the decays into muons and tauons, can undergo significant cancellations for specific values of $a$ and $\alpha$; we have verified this separately, though it is not shown in the figure.

\begin{figure}[t]
    \centering
    \includegraphics[width=0.45\linewidth]{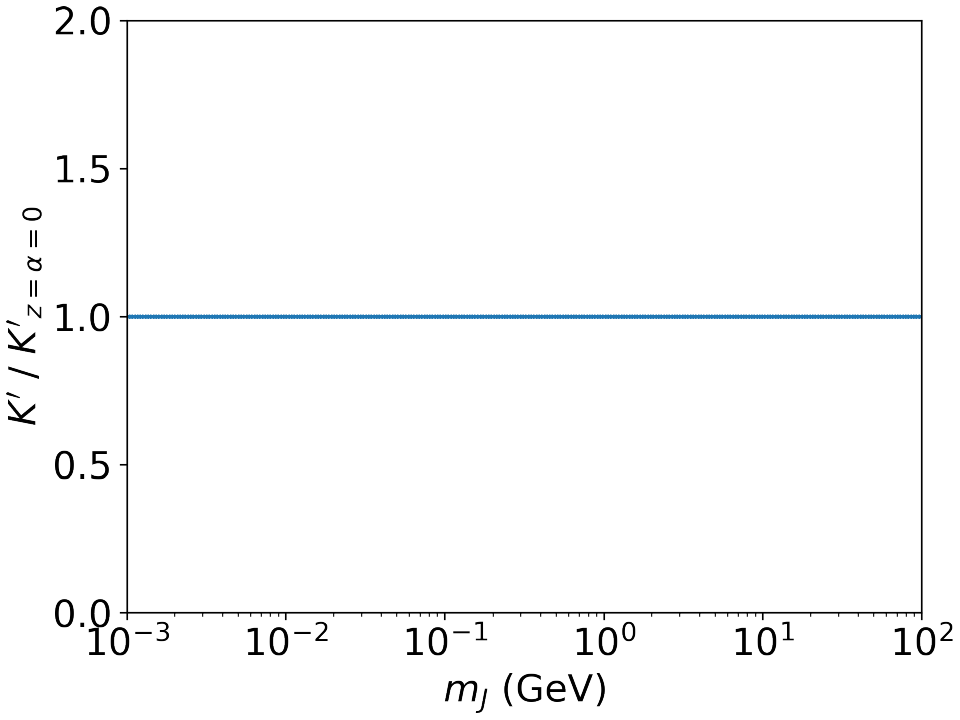}
    \includegraphics[width=0.45\linewidth]{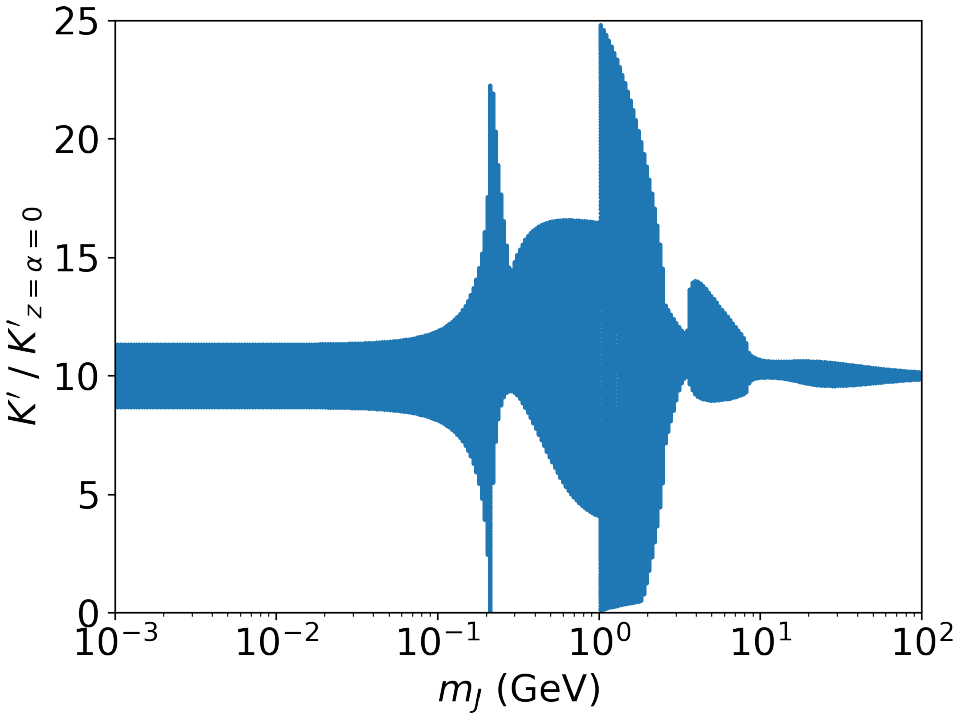}
    \\
    \includegraphics[width=0.45\linewidth]{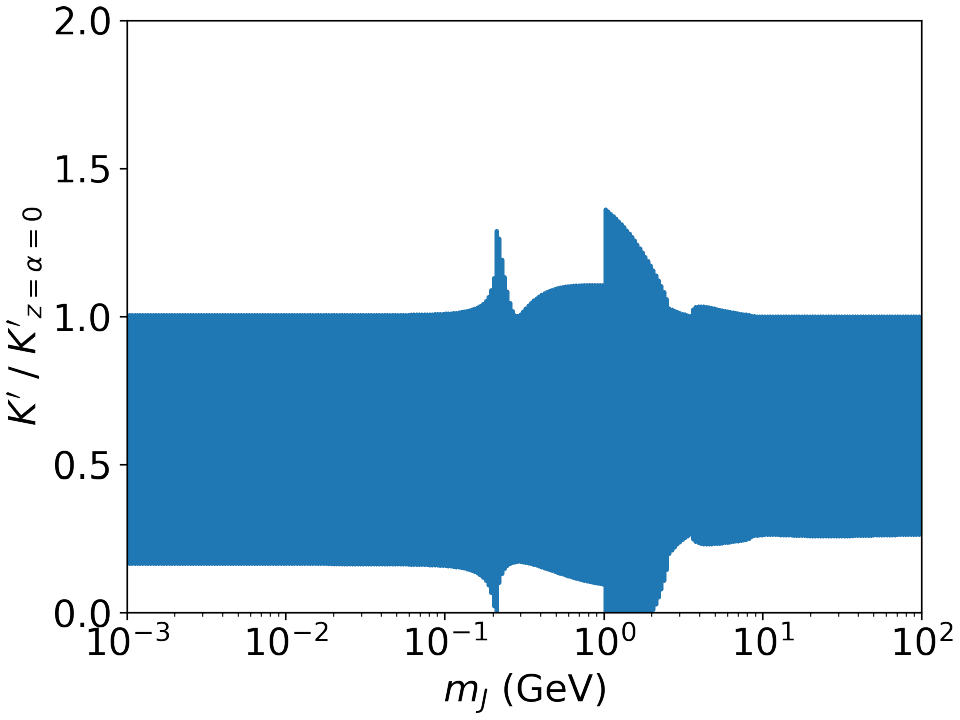}
    \includegraphics[width=0.45\linewidth]{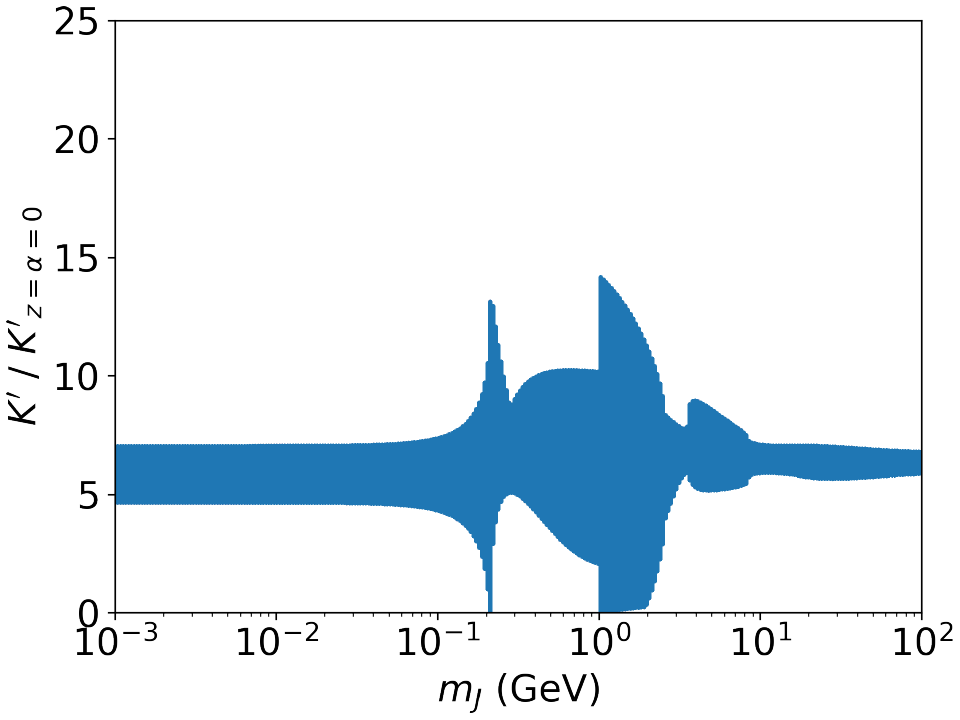}
    \caption{
Dependence of $|K'|$ on the Casas-Ibarra parameter $z=a+ib$ 
and the Majorana phase $\alpha$ as a function of $m_J$, 
for $M_2/M_1=1$ (top) and $M_2/M_1=10$ (bottom), 
with $\cosh 2b=1$ (left) and $\cosh 2b=10$ (right).
The bands show the range obtained by varying $a$ and $\alpha$ 
for each value of $m_J$, normalized to the benchmark case $z=\alpha=0$.
}
\label{fig:Kphoton}
\end{figure}

The dependence of $K'$ (associated with the decay into photons) on $a$ and $\alpha$ varies with $m_J$ due to the function $h(x)$ in Eq.~(\ref{eq:K'}) and (\ref{eq:h(x)}).
Fig.~\ref{fig:Kphoton} shows the dependence of $|K'|$ on $a$ and $\alpha$ for different values of $m_J$, for $M_2/M_1=1, 10$ (top and bottom) and $\cosh{2b}=1,10$ (left and right).
Here, we normalize the value to the $z=\alpha=0$ case for each $m_J$, as in Fig.~\ref{fig:trK-Kee}.
As shown in Fig.~\ref{fig:Kphoton}, $K'$ varies by an $\mathcal{O}(1-10)$ factor across most $m_J$ regions.
However, when $m_J\sim0.2$, $1~\rm{GeV}$, $K'$ can vanish, because of the destructive interference between the terms in Eq~(\ref{eq:K'}) within these $m_J$ intervals.
However, this does not affect the viable parameter regions of Sec.~\ref{sec:production}.

\clearpage

\bibliography{biblio}
\bibliographystyle{JHEP}

\end{document}